\documentclass[onecolumn,showpacs,amsmath,amssymb,aps,showkeys,floatfix,prd,a4paper]{revtex4}

\usepackage[dvips]{graphicx}
\usepackage{dcolumn}
\usepackage{bm}
\usepackage{epsfig}
\usepackage{amsfonts}
\usepackage{amssymb,amscd}

\def\lsim{\raise0.3ex\hbox{$<$\kern-0.75em\raise-1.1ex\hbox{$\sim$}}}
\def\gsim{\raise0.3ex\hbox{$>$\kern-0.75em\raise-1.1ex\hbox{$\sim$}}}

\def\beq{\begin{equation}}
\def\eeq{\end{equation}}
\def\bea{\begin{eqnarray}}
\def\eea{\end{eqnarray}}
\def\bq{\begin{quote}}
\def\eq{\end{quote}}

\def\gappeq{\mathrel{\rlap {\raise.5ex\hbox{$>$}}
{\lower.5ex\hbox{$\sim$}}}}

\def\lappeq{\mathrel{\rlap{\raise.5ex\hbox{$<$}}
{\lower.5ex\hbox{$\sim$}}}}

\def\Toprel#1\over#2{\mathrel{\mathop{#2}\limits^{#1}}}

\def\imag{{\mathcal{I}\mathrm{m}}\,}

\begin{document}


\title{Probing the BFKL dynamics in the Vector Meson Photoproduction at large -- $t$ in  $pPb$ collisions at the CERN LHC}

\author{V.~P. Gon\c{c}alves}
\email{barros@ufpel.edu.br}
\author{W.~K. Sauter}
\email{werner.sauter@ufpel.edu.br}
\affiliation{High and Medium Energy Group \\
Instituto de F\'{\i}sica e Matem\'atica, Universidade Federal de Pelotas\\
Caixa Postal 354, CEP 96010-900, Pelotas, RS, Brazil}
\date{\today}

\begin{abstract}
The photoproduction of vector mesons in $pPb$ collisions at LHC energies is investigated  assuming  that the color singlet $t$-channel exchange carries a large momentum transfer $t$. The rapidity distributions and total cross sections for the process $Pb \, p \rightarrow Pb \otimes V \otimes \,jet + X$, with $V = \rho, \, J/\Psi$ and $\otimes$ representing a rapidity gap in the final state, are estimated considering the non-forward solution of the BFKL equation  at high energy and large -- $t$. A comparison with the predictions obtained at the Born level also is presented. We predict a large enhancement of the cross sections associated to the BFKL dynamics in the kinematical range probed by the LHCb Collaboration. Moreover, our results indicate that the experimental identification can be feasible at the LHC and that this process can be used to probe the BFKL dynamics.

\end{abstract}


\maketitle

The study of hadronic collisions at the  Large Hadron Collider (LHC) offers a unique possibility to probe the theory of the strong interactions -- the Quantum Chromodynamics (QCD) -- in a new and hitherto unexplored energy regime \cite{review_forward}.
 In the last decades, several approaches were proposed to describe the QCD evolution at high energies \cite{hdqcd}. One them is the BFKL approach, proposed by Lipatov and collaborators \cite{BFKL} in the late 1970s to describe the Regge limit of hadronic reactions, which predicts that the cross sections will have a steep increasing with the center -- of -- mass energy. As this behavior implies the violation of unitarity at very high energies, the BFKL approach is expected to be valid only in a limited kinematical window between the standard DGLAP description of the QCD evolution and more sophisticated non -- linear approaches \cite{hdqcd}. Such aspect implies that the probe of the BFKL dynamics is a theoretical and experimental challenge. During the last years,  promissing results have been obtained in treatment of the Mueller -- Navelet jets \cite{wallon,celiberto} and dihadron production separated by a large rapidity gap \cite{celiberto_dihadron} as well  related experimental measurements were performed at the LHC \cite{mn_data}. However, such results indicate e.g. that in order to discriminate between the next - to - leading fixed order calculation and the BFKL prediction for the Mueller -- Navelet jets a very precise measurement would be needed, which become the probe of the BFKL dynamics in these processes a hard task.

An alternative to probe the BFKL dynamics is the study of photon -- induced interactions at the LHC \cite{upc}, where the maximum photon -- photon and photon -- hadron center -- of -- mass energies present in these reactions are at least one order of magnitude larger than those reached in previous $e^+ e^-$ and $e p \,(A)$ colliders. In recent years, several authors have investigated the impact of the BFKL evolution on  observables that can be measured in photon -- induced interactions at the LHC \cite{vicmag_hqgam,vicwer_doublemesons,vicwer1,vicwer2,strikman,ivanov_hq}. Such studies indicated that a future experimental analysis of these observables at the LHC could provide valuable information on the QCD dynamics. In particular, the results presented in Refs. \cite{vicwer1,vicwer2} for the vector meson photoproduction at large momentum transfer in hadronic collisions demonstrated that the magnitude of the cross sections for the photoproduction of $\rho$, $J/\Psi$ and $\Upsilon$ in $pp$ and $AA$ collisions is large and that an experimental analysis is, in principle, feasible at the LHC.
   Our goal in this paper is to complement these previous studies by the analysis of diffractive vector meson photoproduction at large -- $t$ in $pPb$ collisions. Differently from symmetric hadronic collisions ($pp$ and $AA$), where the rapidity distributions of the vector mesons in the final state receive contributions from photon -- hadron interactions with small and large center -- of -- mass energies, in the case of $pPb$ collisions, the process is dominated by $\gamma p$ interactions, with the photons  generated by the nucleus. Consequently, the relation between the rapidity distributions and the energy dependence of  the $\gamma p$ cross section is determined unambiguously. In other words, the analysis of this observable can be used to probe the energy evolution predicted by the BFKL dynamics.

\begin{figure}[t]
\centerline{\psfig{file=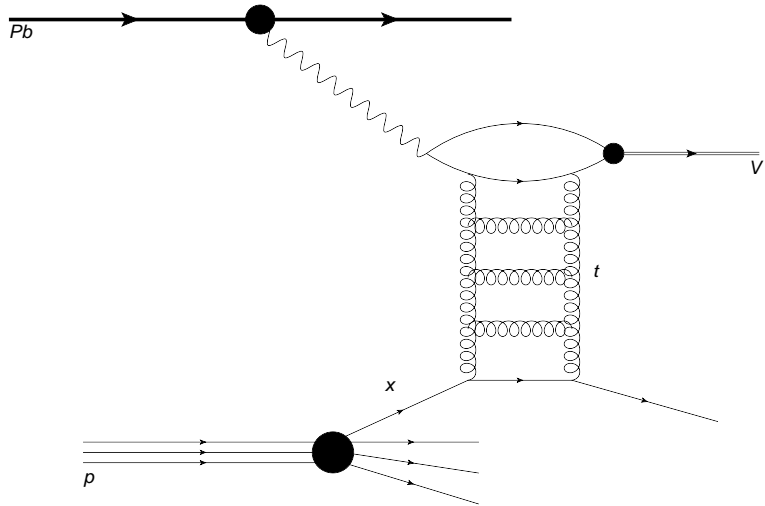,width=100mm}}
 \caption{Vector meson photoproduction at large -- $t$ in $pPb$ collisions.}
\label{fig1}
\end{figure}

 The basic idea in photon -- induced interactions at hadronic collisions is that
the total cross section for a given process at the hadronic level can be factorized in
terms of the equivalent flux of photons into the hadron projectile and
the photon-target production cross section. The main advantage of using colliding hadrons and nuclear beams for studying photon induced interactions is the high equivalent photon energies and luminosities that can be {reached} at existing and
future accelerators (For a review see Ref. \cite{upc}). In the case of the diffractive vector meson photoproduction at large -- $t$  in an ultraperipheral $pPb$ collision, represented by the diagram in Fig. \ref{fig1}, the final state will be characterized by 
two  large rapidity gaps in the detector. One  rapidity gap is predicted to be present between the vector meson and the nuclei, which emits the photon and remains intact. The other gap is expected to be characterized by the  vector meson on one side and a
jet on the other, which balances the transverse momentum. Differently from the exclusive vector meson photoproduction discussed e.g. in Refs. \cite{vicmag_varios,bruno,diego}, in this process  the { $t$-channel color singlet} object carries large momentum transfer, which  means that the square of the four momentum transferred {across} the associated rapidity gap, -$t$, is large, which implies the dissociation of the proton target (For a recent discussion about proton dissociation see e.g. Refs. \cite{wolf,brunodiego}). Consequently, the detection of the final state intact proton or the forward jet can, in principle, be used to discriminate between these two processes.
The cross section for the photoproduction of vector mesons at large -- $t$ in $pPb$ collisions can be expressed by
\begin{eqnarray}
\frac{d\sigma \,\left[Pb + p \rightarrow   Pb \otimes V \otimes jet + X \right]}{dY dt} = \omega \frac{dN_{\gamma/Pb} (\omega )}{d\omega }\,\frac{d \sigma_{\gamma p \rightarrow V \otimes jet +  X}(W_{\gamma p}^2)}{dt}\,
\label{dsigdydt}
\end{eqnarray}
where $\otimes$ characterizes the presence of a rapidity gap in the final state and the rapidity $Y$ of the vector meson produced is directly related to the photon energy $\omega$, i.e. $Y\propto \ln \, (2 \omega/m_{V})$. Consequently, large values of $Y$ are associated to large photon energies. Moreover, $\frac{d\sigma}{dt}$ is the differential cross section for the process $\gamma p \rightarrow V \otimes jet + X $, to be discussed below, and $W_{\gamma p} = [2 \omega \sqrt{s}]^{1/2}$ is the photon -- proton center - of - mass energy, with $\sqrt{s}$ being the c.m.s. energy of the $pPb$ system. { Eq. (\ref{dsigdydt})} implies that, given the nuclear photon flux, the double differential cross {section} is a direct measure of the photoproduction
cross section for a given energy and squared momentum transfer. In particular, the kinematical range of large rapidities probes the $\gamma p$ cross section at large values of $W_{\gamma p}$. Due to coherent condition, the photon flux is characterized by photons with a very low virtuality, which implies that for most purposes, {the photons} can be considered as real. 
Considering the requirement that in photoproduction there is no 
hadronic interaction (ultra-peripheral collision) an  analytical
approximation for the equivalent photon flux of a nucleus can be
calculated, and is given by \cite{upc}
\begin{eqnarray}
\frac{dN_{\gamma/A}\,(\omega)}{d\omega}= \frac{2\,Z^2\alpha_{em}}{\pi\,\omega}\, \left[\bar{\eta}\,K_0\,(\bar{\eta})\, K_1\,(\bar{\eta})- \frac{\bar{\eta}^2}{2}\,{\cal{U}}(\bar{\eta}) \right]\,
\label{fluxint}
\end{eqnarray}
where
  $\gamma_L$ is the Lorentz boost  of the $Pb$ beam; $K_0(\bar{\eta})$ and  $K_1(\bar{\eta})$ are the
modified Bessel functions.
Moreover, for $pPb$ collisions we have that $\bar{\eta}=\omega\,(R_{Pb} + R_{p})/\gamma_L$ and  ${\cal{U}}(\bar{\eta}) = K_1^2\,(\bar{\eta})-  K_0^2\,(\bar{\eta})$, , where $R_p$ and $R_{Pb}$ are the proton and lead radius, respectively. In our calculations we assume that $R_p = 0.7$ fm and $R_A = 1.2 \times A^{\frac{1}{3}}$ fm.
Considering that for $pPb$ collisions at the Run 2 LHC energy the Lorentz factor  is $\gamma_L = 4348$, one obtain that the maximum {c.m.} $\gamma N$ energy is given by $W_{\gamma p} \approx 1300$ GeV. Therefore, while studies at HERA were limited to photon-proton center of mass energies of about 200 GeV, photon-hadron interactions at  LHC can reach one order of magnitude higher on energy. Consequently, studies of photon -- induced  interactions in $pPb$ collisions at the  LHC provide valuable information on the QCD dynamics at high energies.

Let's now discuss the diffractive vector meson photoproduction at large momentum transfer. In this kinematical limit,  the pomeron couples predominantly to individual partons in the proton (For a detailed discussion see, e.g. 
 Refs. \cite{FR,ivanov,BFLW,jhep}). As a consequence,  the cross section for the photon - proton interaction can be expressed by the product of the parton level cross section and the parton distributions of the proton,
\begin{eqnarray}
\frac{d\sigma (\gamma p \rightarrow V \otimes jet + X)}{dt} & = & \int dx_J \, \frac{d\sigma (\gamma p \rightarrow V \otimes jet + X)}{dx_J dt}  \\
& = & \int dx_J \,  \left[ \frac{81}{16} G(x_J,|t|) + \sum_i ( q_i(x_J,|t|) + \bar{q}_i(x_J,|t|))\right] \, \frac{d\sigma}{dt}(\gamma q \rightarrow V \otimes q)\,\,,
\label{dsigdtdx}
\end{eqnarray}
where  $x_J$ is the fraction of the longitudinal momentum of the incoming proton carried by the jet and $G(x_J,|t|)$ and $q_i(x_J,|t|)$ are the gluon and quark distribution functions, respectively. The variable $x_J$ is defined by $x_J = -t/(-t + M_X^2 - m^2)$, where $M_X$ is the mass of the {products of the proton dissociation} and $m$ is the proton mass. The minimum value of $x_J$ is calculated considering the experimental cuts on $M_X$.  The differential cross-section for the $\gamma q \rightarrow V \otimes q$ process, characterized by the invariant collision energy squared $\hat{s} = x_J W_{\gamma p}^2$ of the photon - proton system, is expressed in terms of the amplitude  ${\mathcal{A}}(\hat{s},t)$
as follows
\begin{equation}
\frac{d \sigma}{dt}(\gamma q \rightarrow V \otimes q) = \frac{1}{16 \pi} |{\mathcal{A}}(\hat{s},t)|^2.
\label{dsdt}
\end{equation}
At high energies the amplitude can be factorized in terms of the impact factors of the colliding particles and the Green's function of two interacting reggeized gluons, which is determined by the BFKL equation (For details see e.g. Ref. \cite{jhep}).
At the Born level the process is described by a two gluon exchange and an energy independent $\gamma p$ cross section. At higher orders, the dominant contribution is given by the QCD pomeron singularity which is generated by the ladder diagrams with the (reggeized) gluon exchange along the ladder and is described by the BFKL equation \cite{BFKL}. The exchange of a gluon ladder with interacting gluons generates a cross section that increases with the energy. In order to investigate the impact of the BFKL evolution we will compare, in what follows, the predictions obtained at the Born level with those derived using the non-forward solution of the BFKL equation in the leading logarithmic approximation { (LLA)} and lowest conformal spin \cite{Lipatov} to calculate the scattering amplitude. The BFKL amplitude is given by \cite{vicwer1} 
\begin{equation}
\imag {\mathcal{A}}(\hat{s},t) =
\frac{2}{9\pi} \,\int d\nu \frac{\nu ^{2}}{(\nu ^{2}+1/4)^{2}}e^{\chi (\nu )z}I_{\nu }^{\gamma V}(Q_{\perp })I^{q q}_{\nu }(Q_{\perp })^{\ast },
\end{equation}
 where the quantities $I_{\nu }^{\gamma J/\Psi}$ and $I_{\nu }^{qq}$ are given in terms of the $\gamma \rightarrow V$ and $q \rightarrow q$  impact factors and the BFKL eigenfunctions (For details see Refs. \cite{vicwer1,vicwer2}). The variable  
 $Q_{\perp}$ is the momentum transferred, $t=-Q_{\perp}^2$. Moreover, we have that 
\begin{eqnarray}
z = \frac{3\alpha_{s}}{2\pi} \ln \biggl( \frac{\hat{s}}{\Lambda^{2}} \biggr),
\end{eqnarray} 
and  
\begin{equation}
\chi (\nu )=4{\mathcal{R}}\mathrm{e}\biggl (\psi (1)-\psi \bigg (\frac{1}{2}+i\nu \bigg )\biggr ) \,\,,\label{eq:kernel}
\end{equation}
which is proportional to the BFKL eigenvalues~\cite{Jeff-book} with $\psi(x)$ being the digamma function.

As discussed in previous studies \cite{FP,jhep,vicwer1,vicwer2},  the values of $\alpha_s$ and $\Lambda$ are arbitrary at leading order. As in these analysis, we will assume  that $\Lambda$   depends on a hard scale present in the process and a constant value for $\alpha_s$. In our analysis we will disregard a possible $t$ dependence in the running of $\alpha_s$. As demonstrated in Ref. \cite{jhep}, if this dependence is assumed in $\alpha_s$, the resulting $t$ dependencies of the cross sections are too steep and incompatible with the HERA data. Moreover, it is important to point out that the magnitude of the cross sections is strongly dependent on the value for the coupling constant, being $\propto \alpha_s^4$. Therefore, our predictions are strongly sensitive to the choice of $\alpha_s$. We assume  $\alpha_s = 0.21$, which allows to describe the HERA data. Following Ref. \cite{FP} we {assume} that $\Lambda$ can be expressed by $\Lambda^2 = \beta M_{V}^2 + \gamma |t| $, with $\beta$ and $\gamma$ being free parameters adjusted to describe the HERA data for $\rho$ and $J/\Psi$ production. In order to estimate these parameters we have estimated the  $\gamma p $ cross sections for the $\rho$ and $J/\Psi$ photoproduction at HERA using  the MSTW2008LO parameterization \cite{mstw} for the parton distributions and considering the experimental cuts assumed by the H1 and ZEUS Collaborations. We have obtained that these data can be described by the BFKL solution assuming $\beta = 1$ for both mesons and $\gamma = 0.1 \, (3.0)$ for the $\rho$ ($J/\Psi$) production. We have verified that other combinations of values for $\beta$ and $\gamma$ imply that the $t$ - dependencies of the cross sections are not well described in the region of small or large values of $t$ (See Refs. \cite{vicwer1,vicwer2}). As a consequence, the main parameters present in our LO BFKL analysis have been constrained by the HERA data, which implies that its predictions for the photoproduction at the LHC are, in principle, parameter free. However, some comments are in order before to present our results. Firstly, in our analysis we will disregard the  contribution of higher conformal spins ($n > 0$). As demonstrated e.g. in Ref. \cite{royon}, the contribution of these higher components  become important in the production of Mueller - Navelet jets in hadronic collisions when the rapidity interval of jets is small and the jets have a large transverse energy. In the case of the vector meson photoproduction at large - $t$, the results presented in Ref. \cite{jhep} indicate that the impact of higher conformal spins is small in the range probed by HERA, although the impact increases at large - $t$. These authors have demonstrated that the correction due to the $n \neq 0$ components decreases with increasing the energy $\sqrt{\hat{s}}$ and increases with increasing the ratio $|t|/M_V^2$. Considering the estimates presented in Ref. \cite{jhep} we expect that the impact of the higher conformal spins will be a reduction in our predictions by $\approx 15 \%$ if the  range of large values of $t$ ($10 < -t < 30$) is probed by the LHC. At smaller values of $t$ the impact becomes negligible. Second, our predictions are associated to a leading order calculation. It is well known that next - to - leading logarithmic (NLL) corrections to the BFKL kernel can be large \cite{bfklnlo}. Additionally, in order to implement a full next - to - leading order (NLO) calculation, we also should to take into account the NLO corrections to the $\gamma \rightarrow V$ impact factor \cite{ivanov_nlo}.  The treatment of both these corrections  still is theme of intense debate and its implementation goes beyond the scope of our phenomenological analysis. The main expectation is that the NLO corrections will imply a slower energy dependence of the BFKL cross sections. Results presented in Ref. \cite{jhep}, which have considered part of the NLL corrections to the BFKL kernel, indicate that the LO and NLL predictions are similar in the HERA kinematical range but start to be different at higher energies, in particular in the behavior at large - $t$. As a consequence, our predictions should be considered a upper bound for the magnitude of the rapidity distributions and total cross sections. Surely, a full NLO calculation is important to obtain more reliable predictions for the vector meson photoproduction at LHC energies. However, we believe that the exploratory analysis perfomed in what follows is useful to indicate the potentiality of this process to probe the BFKL dynamics as well as to constrain the influence of the higher order corrections. In particular, we would like to point out that the measurement of the vector meson production at two different rapidities can be used to quantify the energy dependence of the BFKL dynamics and, consequently, to discriminate between different approaches for the treatment of the NLO corrections.

\begin{figure}[t]
\includegraphics[scale=0.5] {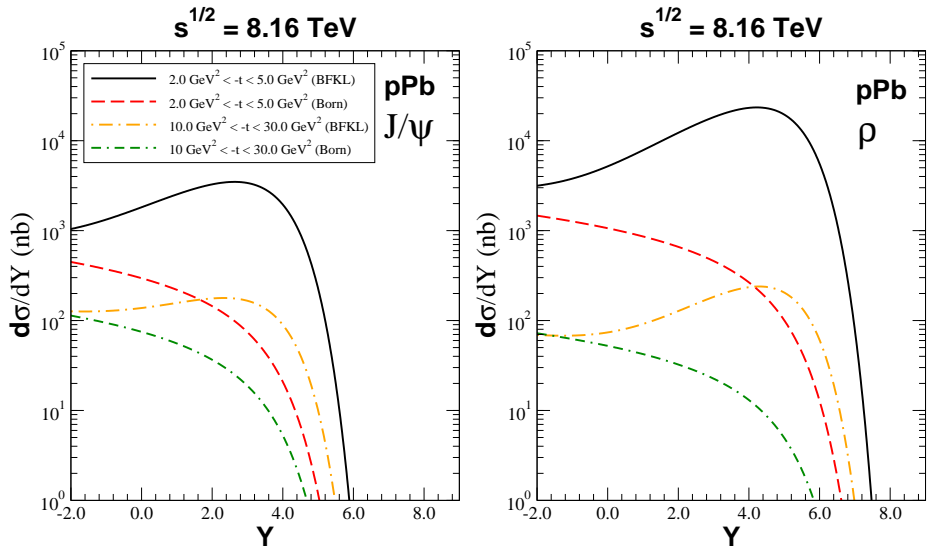}
\caption{ Comparison between the BFKL and Born level predictions for the rapidity distributions for the diffractive $J/\Psi$ (left panel) and $\rho$ (right panel)  photoproduction in $pPb$ collisions at LHC for distinct $t$-ranges and $\sqrt{s} = 8.16$ TeV. }
\label{fig2}
\end{figure}

\begin{figure}[t]
\includegraphics[scale=0.5] {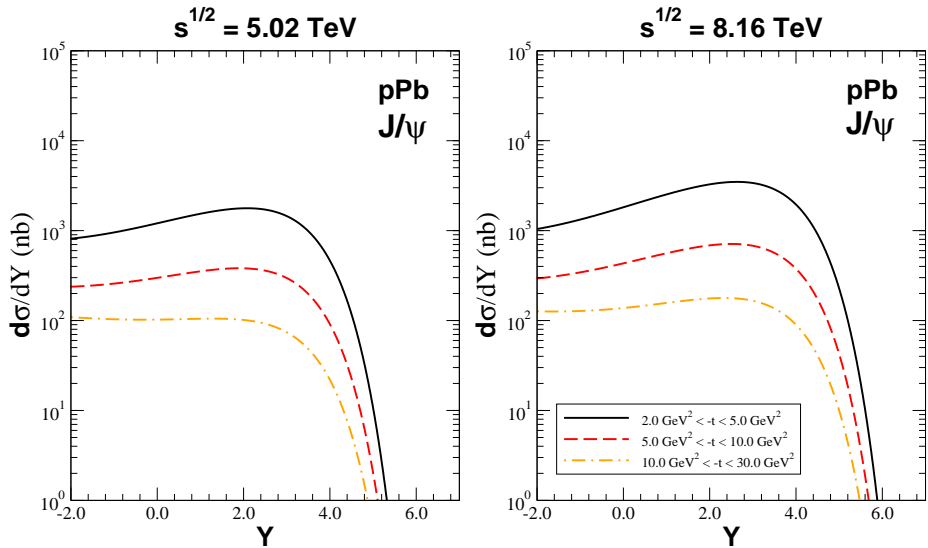}
\caption{Rapidity distributions for the diffractive $J/\Psi$ photoproduction in $pPb$ collisions at LHC energies assuming distinct $t$-ranges and different values of the center-of-mass energy:  $\sqrt{s} = 5.02$ TeV (left panel) and   $\sqrt{s} = 8.16$ TeV (right panel). }
\label{fig3}
\end{figure}

\begin{figure}[t]
\includegraphics[scale=0.5] {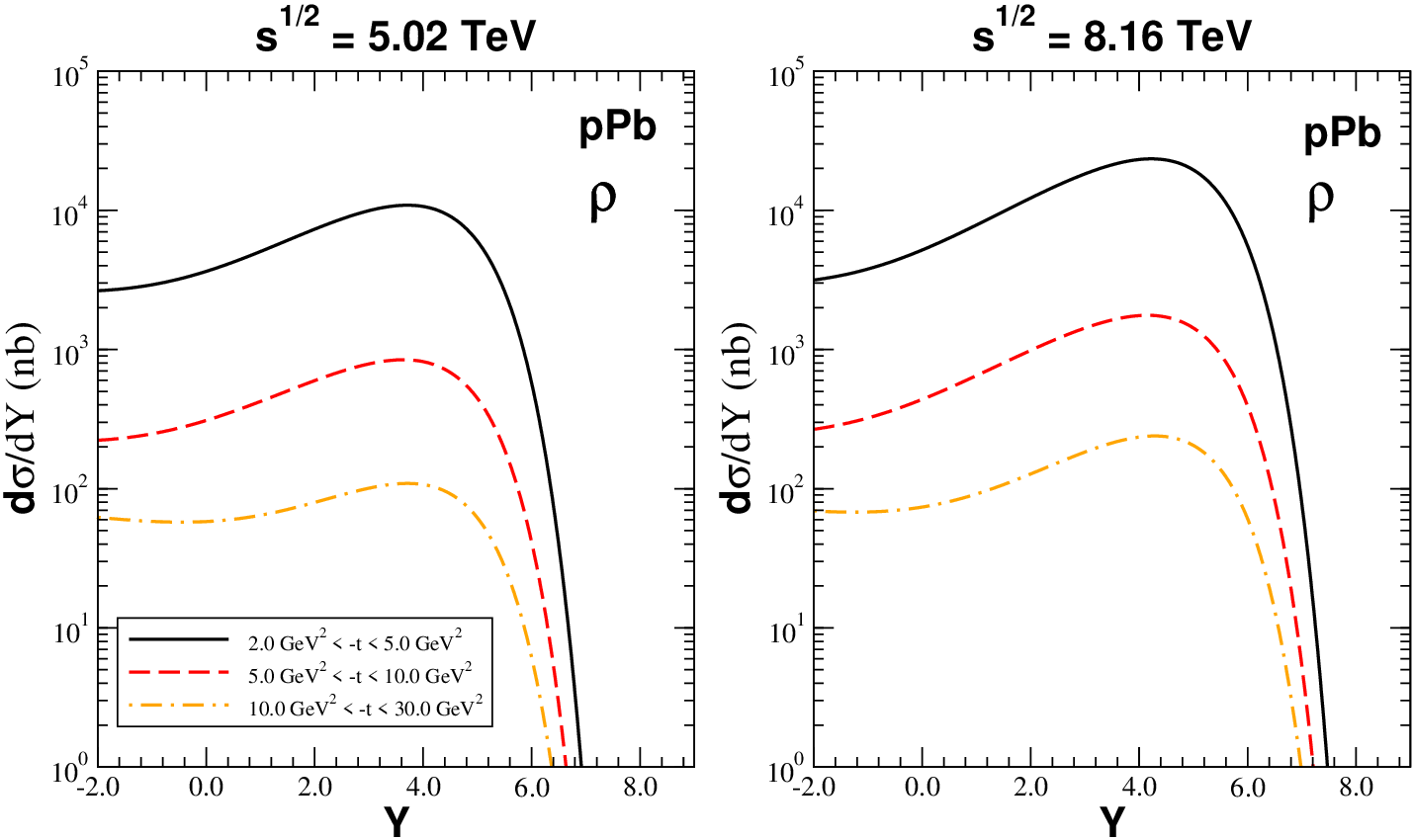}
\caption{Rapidity distributions for the diffractive $\rho$ photoproduction in $pPb$ collisions at LHC energies assuming distinct $t$-ranges and different values of the center-of-mass energy:  $\sqrt{s} = 5.02$ TeV (left panel) and   $\sqrt{s} = 8.16$ TeV (right panel).}
\label{fig4}
\end{figure}

Our predictions for the rapidity distributions for the diffractive $J/\Psi$ and $\rho$ photoproduction in $pPb$ collisions at Run 2 energies are presented in Figs. \ref{fig2}, \ref{fig3} and \ref{fig4}. Following Refs. \cite{FP,jhep} we calculate $d \sigma /dt$ for the {process} $\gamma p \rightarrow V \otimes jet X$ {by} integrating Eq. (\ref{dsigdtdx}) over $x_J$ in the region $0.01 < x_J < 1$. It {means} that we {assume} that the upper limit on the mass of the proton dissociation products at LHC is similar to that considered at HERA. {We have verified that our  predictions} for the rapidity distribution increase by $\approx
$ 20\% if  the minimum value of $x_J$ is assumed to be $10^{-3}$. Moreover, we estimate the Eq. (\ref{dsigdydt}) by integrating over the squared transverse momentum considering 
three different choices for the limits of integration. Basically, we assume the same values used in Ref. \cite{zeus_data} by ZEUS Collaboration. This choice is directly associated to the fact that the associated $\gamma p$ data are quite well described by
our formalism (See Refs. \cite{vicwer1,vicwer2}). 
Initially, in Fig. \ref{fig2},  we present a comparison between the BFKL predictions and those obtained at the Born level, which describe the interaction in terms of a two gluon exchange and implies a $\gamma p$ cross section that is independent of the energy. In order to estimate the impact of the higher orders in the gluonic ladder, associated to the BFKL resummation, we consider in both calculations a common value for the strong coupling constant ($\alpha_s = 0.21$). The predictions for the $J/\Psi$ ($\rho$) are presented in the left (right) panel considering two $t$ -- ranges and a fixed center -- of -- mass energy ($\sqrt{s} = 8.16$ TeV). We have that the BFKL and Born predictions are similar for small values of rapidity and very distinct at large -- $Y$. Such behaviors are expected due to the strict relation between $Y$ and $W_{\gamma p}$. As explained before, at small -- $Y$ we are probing $\sigma_{\gamma p}$ at small values of 
$W_{\gamma p}$, where we expect that BFKL resummation is not effective and its predictions reduce to the Born one. On the other hand, at large -- $Y$ we are probing the behavior of $\sigma_{\gamma p}$ at large energies. In this region, we expect a large impact of the BFKL evolution. Such impact is clearly observed in Fig. \ref{fig2}, where we have that the BFKL and Born predictions differ by a factor $\approx 15$ ($\approx 10$) at $Y = 3$ in the range $2 \le -t \le 5$ GeV$^2$ ($10 \le -t \le 30$ GeV$^2$) in the $J/\Psi$ case. For the $\rho$ production, we predict similar factors at $Y = 3$. In both cases, the enhancement increases at larger rapidities ($Y \approx 4 - 5$). However, at very large rapidities the distributions drop due to the steep decreasing of the photon flux at large values of $\omega$ \cite{upc}. It is important to emphasize that the enhancement associated to the BFKL evolution occurs exactly in the kinematical range probed by the LHCb Collaboration, which implies that our predictions can, in principle, be tested by a future experimental analysis. Another important aspect, already pointed out before, is that if the vector meson production is measured for two different rapidities, we will have access to the energy dependence of the BFKL predictions and, consequently, it will be possible, in principle, to discriminate between the different approaches for the treatment of the NLO corrections.

Our predictions for the diffractive $J/\Psi$ photoproduction in $pPb$ collisions considering two values for the center -- of -- mass energy are presented in Fig. \ref{fig3}. We have that rapidity distributions are smaller when a range with larger values of $t$ is considered; Such behavior is expected from the steep decreasing with $t$ present in $d\sigma/dt$.  Moreover, we have that the predictions increase with the energy, which is directly associated to the BFKL evolution that predicts a strong growth with $W_{\gamma p}$ for the $\gamma p $ cross section ($\sigma_{\gamma p} \propto W^{\lambda}_{\gamma p}$ with $\lambda \approx 1.4$). Similar behaviors are predicted for the diffractive $\rho$ photoproduction, as observed in Fig. \ref{fig4}. The basic differences are the larger values for the rapidities distributions and the effective parameter $\lambda$ that describes the energy dependence of the $\gamma p$ cross section, which is $\approx 1.6$ for the $\rho$ case.

\begin{table}[t]
\begin{center}
\begin{tabular}{||c|r@{\,}l|r@{\,}l|r@{\,}l||}
\hline
\hline
  & \multicolumn{2}{c|}{$pp$ ($\sqrt{s} = 13$ TeV)} & \multicolumn{2}{c|}{$pPb$ ($\sqrt{s} = 5.02$ TeV)} & \multicolumn{2}{c|}{$pPb$ ($\sqrt{s} = 8.16$ TeV)}  \\
\hline
$2.0 < |t| < 5.0$  & 64.9  (507.5) & nb & 12.8  (65.8) & $\mu$b & 20.1  (117.0) & $\mu$b   \\
\hline
$5.0 < |t| < 10.0$  & 12.2  (33.4) & nb & 3.4  (5.3) & $\mu$b & 5.1  (9.2) & $\mu$b \\
\hline
$10.0 < |t| < 30.0$  & 3.2    (6.1) & nb & 1.5  (1.1) & $\mu$b & 1.9  (1.6) & $\mu$b\\
\hline
\hline
\end{tabular}
\end{center}
\caption{The integrated cross sections for the diffractive $J/\Psi$ ($\rho$) photoproduction at large momentum transfer in $pp$ and $pPb$  collisions at Run 2 LHC energies.} 
\label{tab1}
\end{table}

In Table \ref{tab1} we present our predictions for the  total cross sections in $pPb$ collisions at $\sqrt{s} = 5.02$ and 8.16 TeV. For completeness, we also show our predictions for $pp$ collisions at $\sqrt{s} = 13$ TeV, which were not presented in our previous studies \cite{vicwer1,vicwer2}. As the  BFKL dynamics implies a cross section that increases with the energy, we have an enhancement {by} $\approx 70 \%$  when the energy is increased from 5.02 to 8.16 TeV. Moreover, as the photon flux is proportional to $Z^2$, {because} the electromagnetic field surrounding the ion is much larger than the proton one due to the coherent action of all protons in the nucleus,  the nuclear $pPb$ cross sections are amplified by a factor of order of $Z^2$ in comparison to the $pp$ one. As a consequence, we predict large cross sections for the diffractive $J/\Psi$ and $\rho$  photoproduction at large-$t$ in $pPb$ collisions at LHC.


Finally, lets summarize our main results and conclusions. In this paper we investigated the possibility to probe the BFKL dynamics in the photoproduction of vector mesons at large -- $t$ in $pPb$ collisions at the LHC. Our motivation to study this process in $pPb$ collisions is associated to the fact that in this case the photon -- hadron interactions are dominated by photons coming from the nuclei. As a consequence, the analysis of the rapidity distribution gives direct access to the energy dependence of the $\gamma p$ cross section. Therefore, it allows to study the impact of the BFKL evolution on the $t$-- channel color singlet gluonic object present in the diffractive interaction. At the Born level, it is described by a two gluon system and the $\gamma p$ cross section is energy independent. On the other hand, the BFKL approach predicts a steep behavior. Our results demonstrated that this behavior can be probed by the analysis of the rapidity distributions for the diffractive $\rho$ and $J/\Psi$ photoproduction in $pPb$ collisions at Run 2 energies. In particular, we predict large values for the total cross sections and that  the enhancement in the rapidity distributions, associated to the BFKL evolution, should to occur in the kinematical range probed by the LHCb Collaboration. Our results indicate that a future experimental analysis at the LHC is, in principle, feasible and that this process can be used to constrain some aspects of the QCD dynamics at high energies.

\section*{Acknowledgements}
 VPG would like to thanks R. McNulty, M. Rangel and W. Schafer by useful discussions. VPG is grateful to the members of the THEP group for the hospitality at  Lund University, where the revised version of this work was finished. This work was partially financed by the Brazilian funding agencies  CNPq,  FAPERGS and INCT-FNA (process number 464898/2014-5).



\end{document}